\documentclass[12pt]{cernrep}

\usepackage{cite}
\usepackage{epsfig}
\usepackage{graphics}
\usepackage{inputenc}
\inputencoding{latin1}

\begin{document}

\sloppy

\begin{titlepage}
{\par\raggedleft \texttt{MAN/HEP/2005/3}~\\
\texttt{24 August 2005}\par}
\bigskip{}

\bigskip{}
{\par\centering \textbf{\large Central Exclusive Di-jet Production at the Tevatron}\large \par}
\bigskip{}

{\par\centering B. E. Cox and A. Pilkington \\
\par}
\bigskip{}
{\par \centering \small {School of Physics and Astronomy, University of Manchester }\\
 {\small Manchester M13 9PL, England} \par}
\bigskip{}

\begin{abstract}
\noindent

We perform a phenomenological analysis of dijet production in double pomeron exchange at the Tevatron. We find that the CDF Run I results do not rule out the 
presence of an exclusive dijet component, as predicted by Khoze, Martin and Ryskin (KMR). With the high statistics CDF Run II data, we predict that an exclusive component 
at the level predicted by KMR may be visible, although the observation will depend on accurate modelling of the inclusive double pomeron 
exchange process. We also compare to the predictions of the DPEMC Monte Carlo, which contains a non-perturbative model for the central exclusive process. We show that the 
perturbative model of KMR gives different predictions for the di-jet $E_T$ dependence in the high di-jet mass fraction region than non-perturbative models.      

\end{abstract}

\end{titlepage}

\section{Introduction}
\label{sec:intro}

The detection of new particles in central exclusive production at the LHC has received a great deal of attention recently 
(see for example \cite{Cox:2001uq, DeRoeck:2002hk,Boonekamp:2004nu,Assamagan:2004mu,Cox:2004rv,Khoze:2005ie} and references therein).
By central exclusive production, we refer to the process ${ pp \rightarrow p~\oplus~\phi~\oplus~p}$, where $\oplus$ denotes the absence of hadronic activity (`gap') 
between the outgoing protons and the decay products of the central system $\phi$, shown schematically in Fig. 1 (a). An example would be Standard Model Higgs boson production, 
for which the central system 
could consist of either 2 b-quark jets or the decay products of two W bosons, and no other activity. Khoze et al. calculate the cross section 
for the central exclusive production 
of a 120 GeV Standard Model Higgs boson to be $3$ fb at the LHC, falling to $\sim 1$ fb at $M_H = 200$ GeV \cite{Khoze:2001xm,Cox:2005if}, and orders of magnitude 
larger for certain MSSM scenarios \cite{Kaidalov:2003ys}. This is large enough to be observable, and there is currently 
 a proposal to install forward proton detectors as a future upgrade to the LHC experiments \cite{Albrow:2005ig}. It is therefore 
important to confront the KMR calculations with data as soon as possible, in particular by searching for high-rate central exclusive processes at the Tevatron. 

There are several exclusive processes which, 
according to the KMR calculations, should have large enough cross sections to be observable now. The two cleanest signatures are central exclusive $\chi_C$ meson production and 
central exclusive di-photon production. These processes are under study by both the CDF and D\O\ Collaborations, although at the time of writing, no results have been published. 
By far the highest rate process 
is predicted to be central exclusive di-jet production.  
Di-jet production in the inclusive process ${ pp \rightarrow p~\oplus~X~\oplus~p}$, where $X$ can be any centrally produced system, was studied by the CDF Collaboration in the Tevatron Run I data \cite{cdfdijet}. This process, often termed double pomeron exchange, 
is conventionally modelled as shown in figure 1(b). In Regge-inspired language, it can 
be pictured as the emission of two pomerons (or reggeons if the longitudinal momentum loss of the protons is large) 
which collide to produce the central system $X$. Since the pomerons have a 
partonic structure, described by structure functions, there are necessarily always pomeron 
remnants; that is to say there are no truly exclusive events in this picture.    
\begin{figure}[htb]        
\label{fig1}
\begin{center}
  \setlength{\unitlength}{1 mm}      
  \Large 
\begin{picture}(150,70)(0,0)
    \put(0,0){\epsfig{file=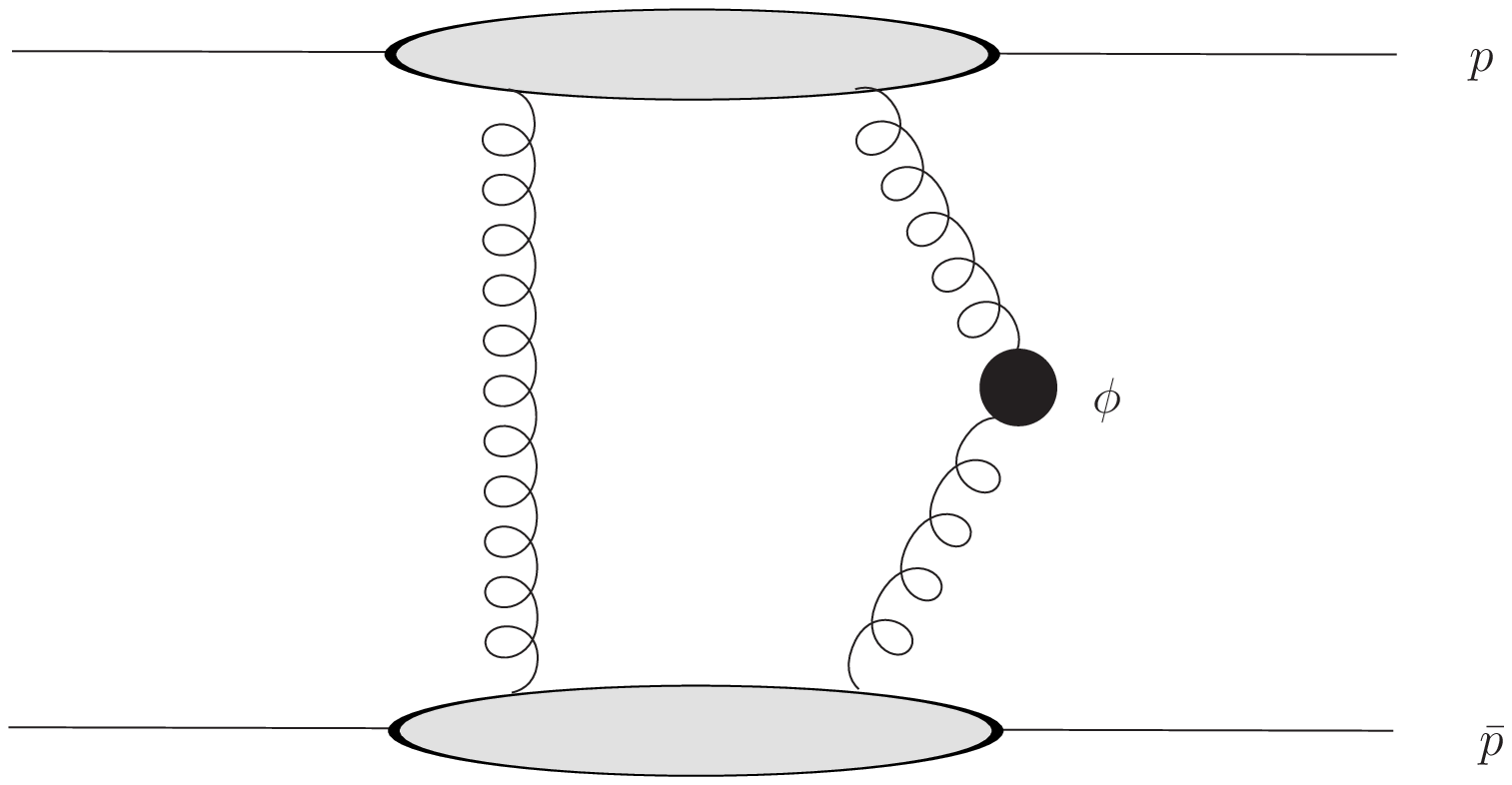,width=0.5\textwidth,height=7cm}}
    \put(75,0){\epsfig{file=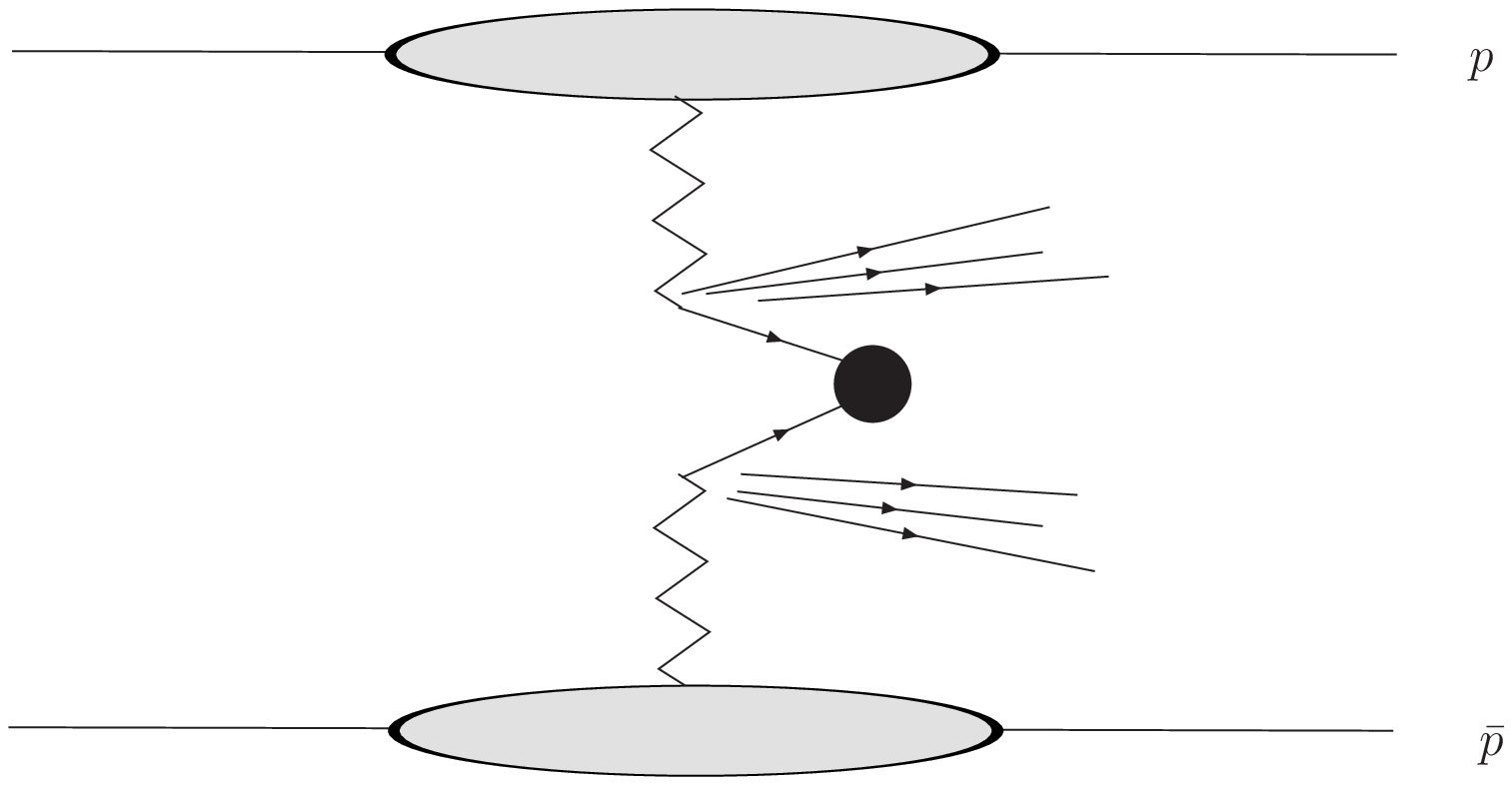,width=.5\textwidth,height=7cm}}
    
\end{picture}
\end{center}
\caption 
        {The exclusive production process (a) and central inclusive (or double pomeron) process (b).}
\end{figure}

Whilst from a theoretical perspective, the exclusive process is distinctly different from the inclusive, from an experimental viewpoint the definition of 
exclusive and inclusive is
somewhat arbitrary.  The approach adopted by the CDF collaboration is to define the quantity $R_{JJ}$ as the fraction of the invariant mass of the central system 
contained within the two highest $E_T$ jet cones. If the cone radius is large enough such that out of cone effects are small, then $R_{JJ}$ should be peaked at or near $1$ for exclusive events since there will be no pomeron remnants. If the exclusive process is present with a large enough cross section relative to double pomeron 
exchange then an excess of events over that predicted by the Regge models should be observable at high $R_{JJ}$.    

In this paper, we carry out a Monte Carlo simulation of di-jet production in double pomeron exchange, and compare directly to the Run I data. 
We use the POMWIG Monte Carlo generator to simulate the inclusive double pomeron exchange process \cite{Cox:2000jt}, and ExHuME \cite{Monk:2005ji} to simulate the exclusive contribution, as calculated by KMR. 
An alternative model for the exclusive process, based on the calculations of Bialas and Landshoff (BL) \cite{Bialas:1991wj}, is implemented in the DPEMC Monte Carlo \cite{Boonekamp:2003ie}. 
There are two key differences between the BL and the KMR approaches. Firstly, in the KMR model, the two gluons at the proton vertex couple perturbatively to the off-diagonal unintegrated gluon 
distribution of the proton. In the BL approach, the proton vertex is treated as non-perturbative, and parameterised using a Regge-motivated ansatz. This leads to a significant difference in the rapidity distributions of the central system; in the BL approach the rapidity distribution is much flatter, since the cross section has only a weak $x$-dependence, entering via the soft pomeron 
intercept, whilst in the KMR model there is a much stronger $x$-dependence from the gluon distributions in the protons. Secondly, the cross section falls much more quickly as the di-jet $E_T$ increases in the KMR model. This is a reflection of the fact that the phase space for gluon emission into the gap increases 
as $E_T$ increases, and therefore the exclusive cross section decreases. This effect is not present in the BL approach. As we shall see, this phase space effect has observable 
consequences which could allow the two models to be separated in the Tevatron Run II data.  

CDF have also recently presented high statistics 
preliminary results on diffractive di-jet production in Run II \cite{Gallinaro:2005qh,Goulianos:2005im}. Since the Run II results 
are only preliminary at the time of writing, we do not compare directly, but make a prediction based on the kinematic range quoted in \cite{Gallinaro:2005qh}.  Our aim is to assess whether 
the exclusive process should be visible in the CDF data.     

\section{Monte Carlo simulation}

To simulate the exclusive process, we use both the ExHuME 1.3.1 generator 
\cite{Monk:2005ji}, interfaced to Pythia 6.205 \cite{Sjostrand:2000wi,Sjostrand:2003wg} for 
parton showering and hadronisation, and DPEMC 2.4 \cite{Boonekamp:2003ie}. To simulate the double pomeron exchange process, we use POMWIG 1.3 \cite{Cox:2000jt}. POMWIG simulates diffractive collisions 
using a Regge factorisation ansatz, with the parameters of the pomeron and reggeon flux terms and structure functions extracted from diffractive DIS data by the 
H1 Collaboration at HERA \cite{Adloff:1997sc}.  
The POMWIG model generates only the inclusive process of figure 1(b) - i.e. there will always be 
pomeron remnants in the event. 
We use the default POMWIG parameters as described in \cite{Cox:2000jt}, with one exception. The default POMWIG behaviour up to v1.2 was to treat the pomeron (and reggeon) 
remnants in the same way that HERWIG treats the photon remnants, i.e. the valence partons in the pomeron and reggeon are defined to be quark - antiquark pairs. 
POMWIG v1.3 introduces the option to define 
the valence partons in the pomeron to be gluons. It is this option that we take to be the default behaviour for the pomeron in this paper. For the Reggeon, we define the 
valence partons to be quarks, as in version 1.2. We discuss the effects of the different remnant treatments in section 4.       
Pomeron and reggeon exchanges are generated separately, and added according to their generated cross sections. POMWIG does not include a model of gap survival probability, 
and therefore over-estimates diffractive cross sections at the Tevatron and LHC.
The approach we adopt is to scale the final POMWIG prediction,
after the experimental cuts and smearing described in section \ref{results},  
to the CDF Run I measurement \cite{cdfdijet}. This assumes that the gap survival factor is independent of kinematic effects, 
which is probably a reasonable approximation given the level of accuracy we are working to here.  
In the exclusive case, the gap survival factor is a prediction of the KMR model. We use the ExHuME default value of  $\mathcal{S}^{2}$ = 0.045 at the Tevatron for both ExHuME and DPEMC 
\footnote{The authors of DPEMC suggest that the gap survival factor to be used in DPEMC 
at the Tevatron is 0.1, although the uncertainty in the normalisation of the DPEMC cross section is at least a factor of 2, which covers the lower value we use here \cite{Boonekamp}.}. 
The normalisation of the ExHuME cross section is particularly sensitive to the gluon distribution in the proton (which enters roughly speaking to the fourth power). 
We use the MRST2002 parton distribution functions \cite{Thorne:2002mr} as the ExHuME default. To gain some estimate of the uncertainty, we also use the CTEQ6M pdfs \cite{Pumplin:2002vw} in our Run II predictions, which lead to a larger cross section prediction \footnote{For a recent analysis of the effect of using different parton distribution functions on the central exclusive cross section, see \cite{HERALHC}}.   
\section{Results}
\label{results}
\subsection{CDF Run I results}
\label{cdf1results}
We first turn our attention to the published CDF Run I results \cite{cdfdijet}. 
CDF have proton tagging detectors only on the outgoing anti-proton side of the experiment. Tagging the outgoing $\bar p$ restricts its longitudinal momentum loss to the range $0.035 < \xi_{\bar p} < 0.095$, 
and the transverse momentum $|t_{\bar p}| < 1$ GeV$^2$. On the outgoing proton side, CDF have no proton taggers, 
so the analysis relies on the observation of a rapidity gap (i.e. no particles above 
a noise threshold) in the forward detectors, covering the range $3.2 < |\eta| < 5.9$. 
This corresponds to an approximate range of proton longitudinal momentum loss 
$0.01 < \xi_{p} < 0.03$. Following the CDF analysis, jets are found using the cone algorithm with cone radius $R=0.7$.
In table 1 we show the generated cross sections from POMWIG, ExHuME and DPEMC in the CDF Run I kinematic range \cite{cdfdijet}, defined as follows;
\begin{eqnarray}
E_T^{jets 1,2} > 7 {\rm GeV} \\
-4.2 < \eta_{jets 1,2}< 2.4 \\ 
0.035 < \xi_{\bar p} < 0.095 \\
|t_{\bar p}| < 1 {\rm GeV}^2 \\
0.01 < \xi_p < 0.03. 
\end{eqnarray}
We note that, at the hadron level, POMWIG overshoots the measured cross section by a factor of $\sim 8$. This would imply an effective gap survival factor of 0.14. As we shall see below however, 
a combination of the smearing of the Monte Carlo, and the rapidity gap definition used by CDF to calculate the measured cross section, changes this result by a factor of $\sim 2$. 
\begin{table}[htdp]
\begin{center}
\begin{tabular}{| c | c | c | c |}
\hline
& $E_{T, min}$ (GeV)   & $\sigma_{TOT}$ (nb) & $\sigma_{R_{JJ} > 0.8}$ (nb) \\
\hline 
CDF & 7 & 43.6$\pm$4.4$\pm$21.6 & $<$ 3.7 \\
POMWIG & 7 & 320.74 & 0.69  \\ 
ExHuME & 7 &   0.48 & 0.08 \\
DPEMC & 7 & 1.32 & 0.26 \\
\hline
CDF & 10 & 3.4$\pm$1.0$\pm$2.0 & n/a \\
POMWIG & 10 & 49.74 &  0.36\\ 
ExHuME & 10 &  0.25 & 0.07 \\
DPEMC & 10 & 0.68 &  0.22\\
\hline
\end{tabular}
\end{center}
\label{tresults}
\caption{Generator Level Cross Sections and CDF Data for Run 1. The POMWIG result has no effective survival factor applied as this is calculated using the smeared data. $R_{JJ}$ is calculated by using the generated $\xi_{1,2}$ values and unsmeared final state hadrons.}
\end{table}%

The CDF Collaboration results are not fully corrected for detector effects. 
In order to compare directly to the published CDF distributions
we must therefore smear the energy and momenta of the final state particles from the Monte Carlos. This is done 
according to the figures quoted in the CDF technical design reports \cite{Bertolucci:1987zn,Blair:1996kx} before jet finding. 

The first step in the CDF analysis is to define a single diffractive di-jet data set, using the proton tagging detector on the outgoing anti-proton 
side, in the kinematic range defined by cuts (1), (2), (3) and (4) above. We smear the energy and momenta of the final state particles, 
but take the $\xi_{\bar p}$ and $t_{\bar p}$ values from the Monte Carlo unsmeared. A calorimeter noise suppression cut is applied at $E_{cell} < 100$ MeV. We apply 
this cut to the smeared energies of the final state particles. 

A rapidity gap on the outgoing proton side is required by demanding no particle hits in the Beam-Beam Counters 
(BBC), covering the pseudorapidity region $3.2 < \eta < 5.9$, and no hits in the Forward Calorimeter towers (FCAL), $2.4 < |\eta| < 4.2$, with $E_{cell} > 1.5$ GeV. This cut approximates to the $\xi_p$ range given in (5). 
We apply the above $\eta$ cuts and calorimeter noise thresholds to the smeared final state Monte Carlo particles. The value of $\xi_p$ is then calculated from the calorimeter information by

\begin{eqnarray}
\xi_{p} = \frac{1}{\sqrt s} \sum_{particles} E_T e^{\eta}
\end{eqnarray}

where the sum is over all final state particles excluding the outgoing protons. The 
calorimeter-based $\xi_p$ 
measurement is calibrated by comparing the direct $ \xi_{\bar p}$ measurement from the proton tagging detector on the outgoing anti-proton side with the corresponding calorimeter-based measurement of 
$ \xi_{\bar p}$,
\begin{eqnarray}      
\xi_{\bar p} = \frac{1}{\sqrt s} \sum_{particles} E_T e^{-\eta}
\end{eqnarray} 
This method allows CDF to calculate an average correction factor which can be used to correct the calorimeter-based $\xi_p$ measurement. CDF find that the $ \xi_{\bar p}$ value 
measured in the calorimeter should be multiplied by a correction factor of 1.7. We have performed an identical procedure on the smeared Monte Carlo and found a smaller correction (1.2) is required. 

This could be due to the fact that CDF had very low statistics in the data sample from which the correction factor was derived, or could indicate that extra detector effects, 
apart from smearing, are present. It could also of course be due to inaccuracies in the POMWIG model. We use our correction factor of 1.2 in the remainder of the Run I analysis - i.e. we make a correction for the kinematical reconstruction method of $ \xi_{p}$, 
but allow for the possibility that additional effects might have been corrected for in the CDF data.    

In figure 2 we compare the $\xi_p$ and $\xi_{\bar p}$ distributions from POMWIG and ExHuME to the CDF Run I data. 
The agreement is reasonable in $\xi_{\bar p}$, which is measured directly in the 
proton tagger. For the calorimeter-based $\xi_p$ distribution, we reproduce the overall shape of the data. We attribute the 
differences in certain bins to statistical fluctuations in the data, or to detector effects that we are unable to 
reproduce, since
one would expect this distribution to be smooth, except in the lowest $\xi$ region, where kinematic factors 
cause the distribution to turn over. In figure 3 we show the mean jet $E_T$ and mean jet $\eta$ distributions of the two highest $E_T$ jets compared to the CDF Run 1 data. We obtain reasonable agreement. 
We attribute the undershooting of the data in the lowest $E_T$ bin, and the more peaked mean $\eta$ distribution in the Monte Carlo, to detector effects. The DPEMC samples (not shown) give similar results. 
The distributions are reasonably well described by POMWIG alone. This was also the conclusion of a similar analysis presented in \cite{Appleby:2001xk}.
\begin{figure}[htb]        
\label{xpom}
\begin{center}
  \setlength{\unitlength}{1 mm}      
  \Large 
\begin{picture}(160,70)(0,0)
    \put(0,0){\epsfig{file=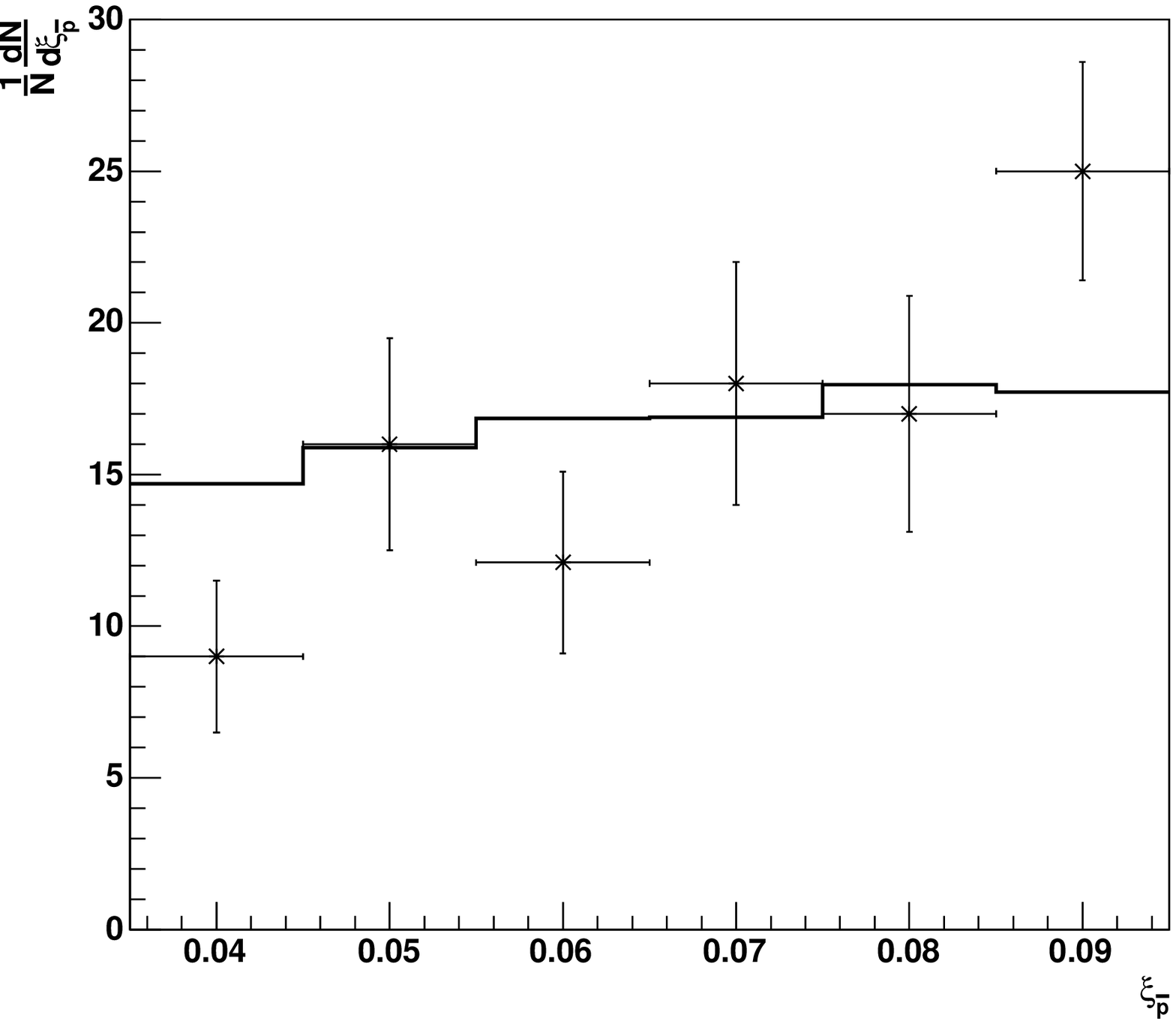,width=.5\textwidth}}
    \put(80,0){\epsfig{file=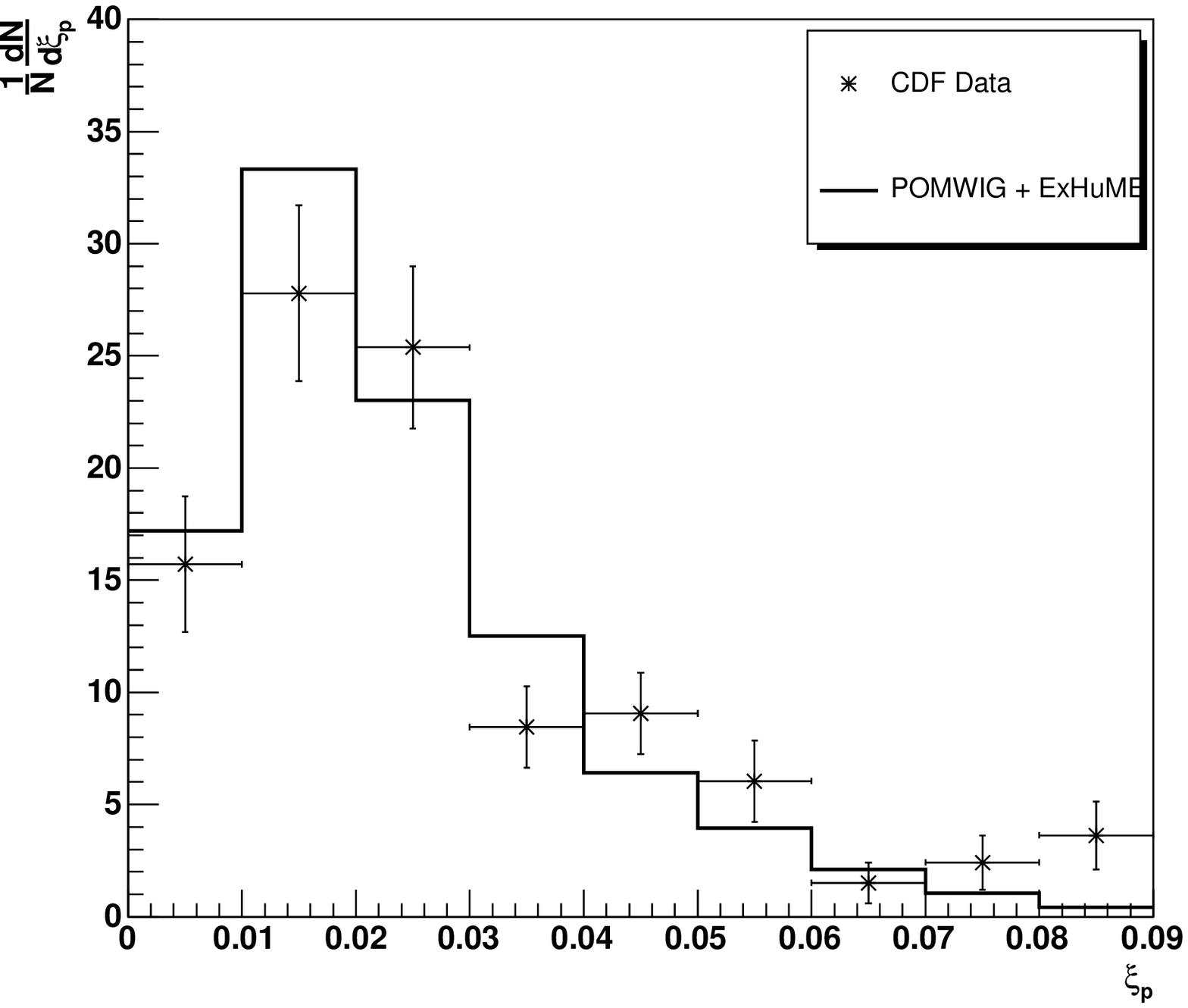,width=.5\textwidth}}       
\end{picture}
\end{center}
\caption 
        {The $\xi_{\bar p}$ and $\xi_p$ distributions compared to CDF Run I data \cite{cdfdijet}. $\xi_{\bar p}$ is measured directly 
in the forward proton tagging detector, and $\xi_p$ is measured using the central detectors, as described in the text. The solid line the sum of the ExHuME and POMWIG predictions.}
\end{figure} 
  
To calculate the cross section, CDF restrict the DPE sample to the range $0.01 < \xi_p < 0.03$. The double pomeron exchange cross section is obtained by taking the ratio of the number of double 
diffractive di-jet to single diffractive di-jet events, and multiplying by the single diffractive di-jet cross section. Using this method, CDF measure the DPE di-jet cross section, in the kinematic 
range defined by cuts (1) - (5), to be   $\sigma^{\rm DPE} = 43.6 \pm 4.4 (stat) \pm 21.6(syst)$ nb.

With the DPE cross section defined in the above kinematic range (i.e. cutting on the generated values of $\xi_{\bar p}$, the scaled calorimeter measurement of $\xi_p$, and the smeared final state 
particles in the jet finding), we find that POMWIG over-estimates the measured cross section by a factor of $\sim 4$, leading to an effective gap survival factor of 0.27. This fixes the normalisation 
of POMWIG. The discrepancy of a factor of $\sim 2$ between the hadron and detector level normalisation factors occurs because of the rapidity gap requirement, which results in the loss of events in 
the $0.01 < \xi_p < 0.03$ range and not just at high $\xi_p$. We note that POMWIG does not simulate reggeon - pomeron collisions, and may therefore underestimate the 
cross section at the Tevatron as predicted from the HERA data. The gap survival factor should therefore be considered as an 'effective' gap survival 
factor which takes into account such effects. It is also worth noting that there is a systematic error of $\sim 50 \%$ on the overall normalisation of the CDF cross section.

\begin{figure}[htdb]        
\label{meaneteta}
\begin{center}
  \setlength{\unitlength}{1 mm}      
  \Large 
\begin{picture}(160,70)(0,0)
    \put(0,0){\epsfig{file=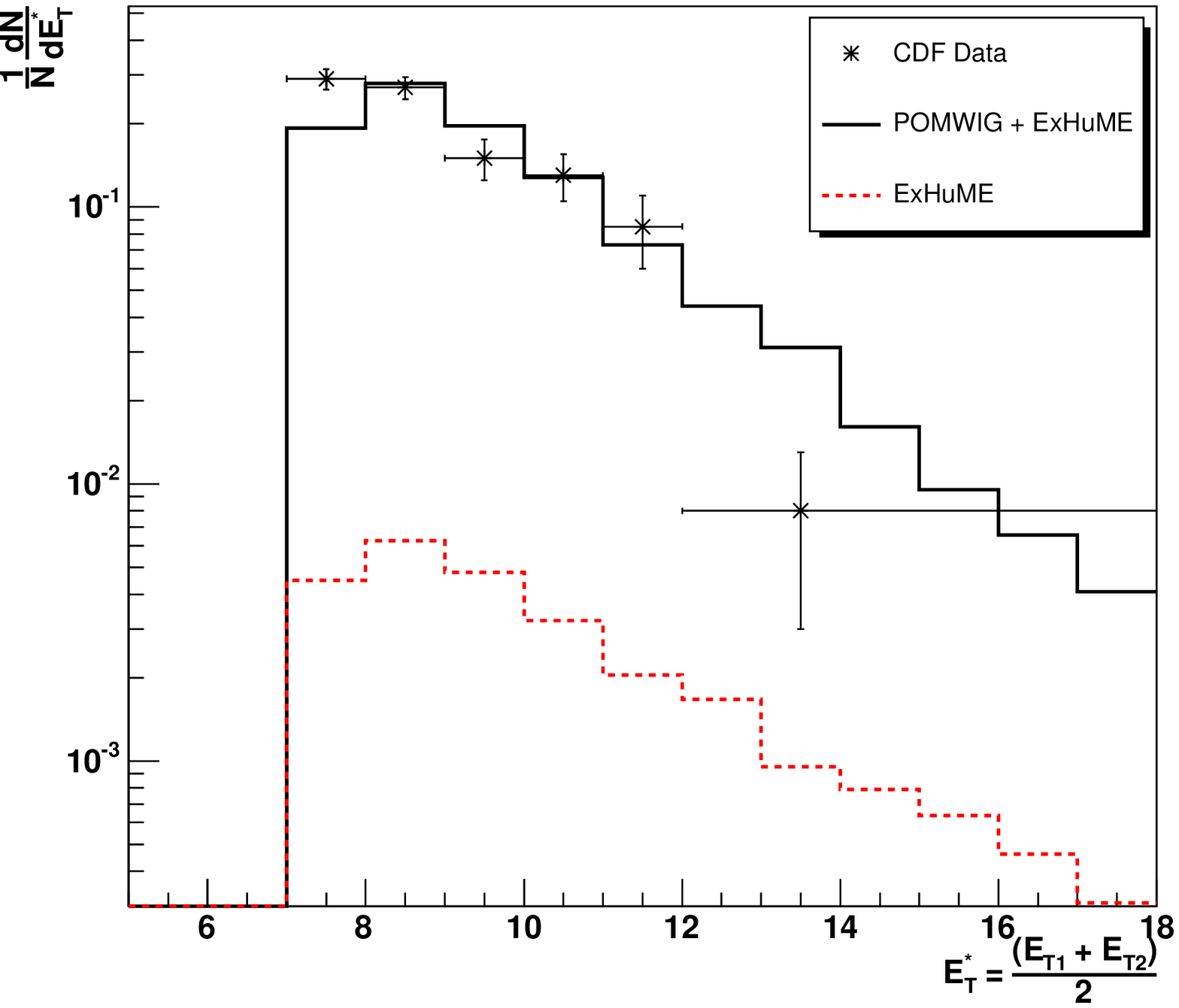,width=.5\textwidth}}
    \put(80,0){\epsfig{file=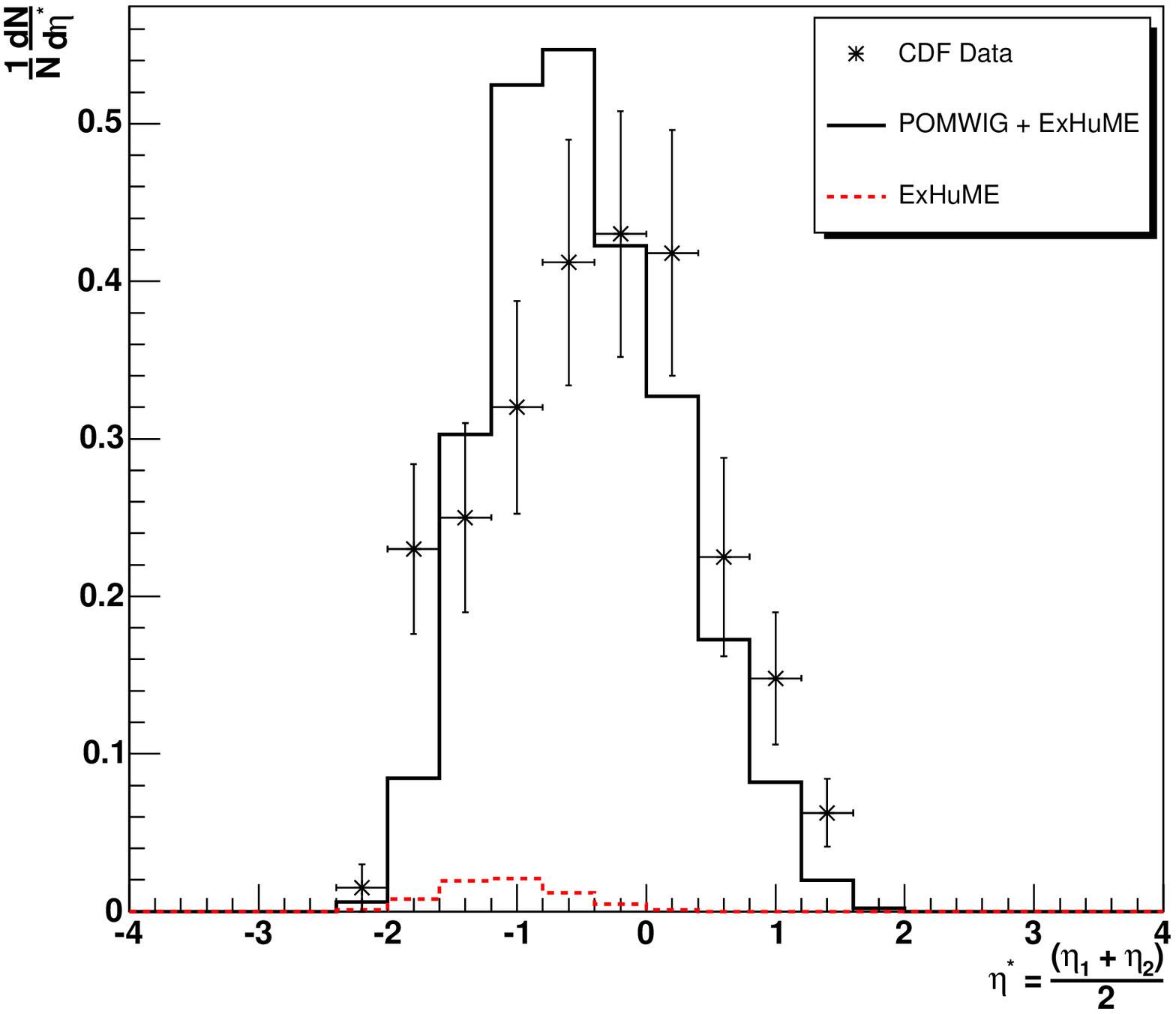,width=.5\textwidth}}
\end{picture}
\end{center}
\caption 
        {The mean jet $E_T$ and mean jet $\eta$ distributions of the 2 highest $E_T$ jets, defined as described in the text.
 The data points are the CDF Run I results \cite{cdfdijet}.  The dashed line is the prediction of the ExHuME Monte Carlo, and the solid line the sum of the ExHuME and POMWIG predictions.}
\end{figure}

\begin{figure}[htb]        
\label{RJJRunI}
\begin{center}
  \setlength{\unitlength}{1 mm}      
  \Large 
\begin{picture}(160,70)(0,0)
    \put(0,0){\epsfig{file=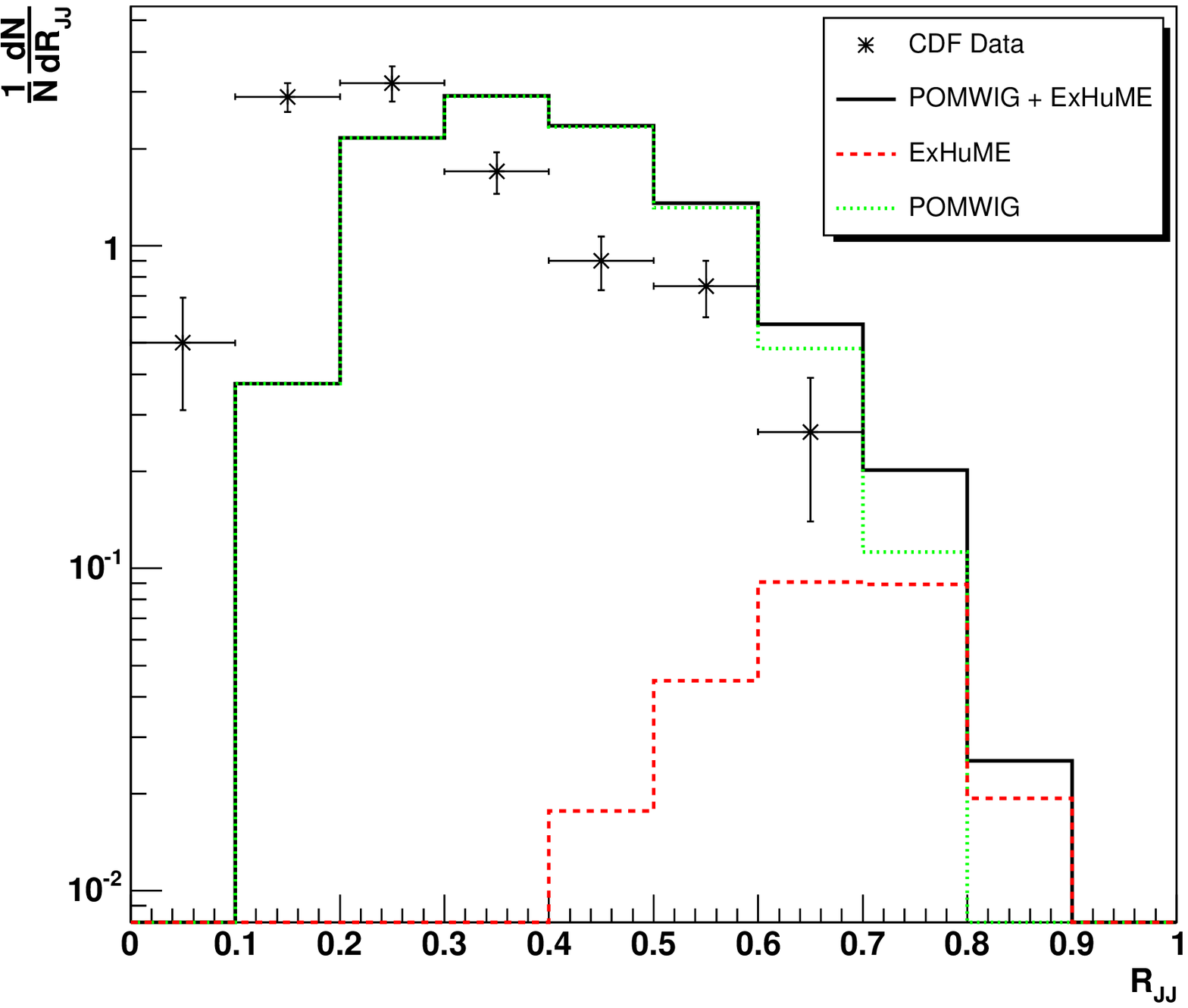,width=.5\textwidth}}
    \put(80,0){\epsfig{file=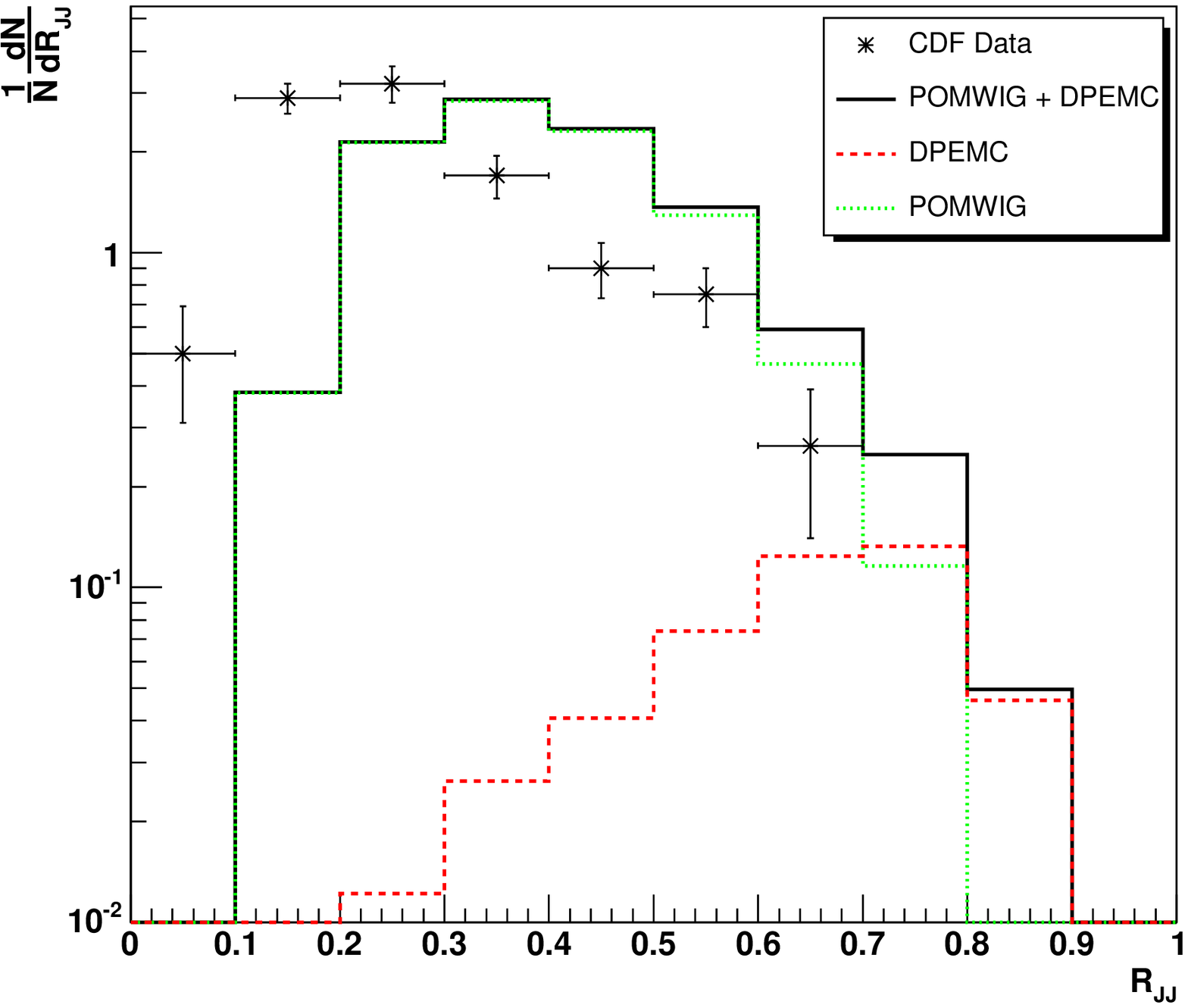,width=.5\textwidth}}
\end{picture}
\end{center}
\caption 
        {The fraction of the invariant mass of the central system contained within the 2 highest $E_T$ jet cones. The dashed line is the 
prediction from ExHuME (i.e. the central exclusive process), the dotted line is the prediction from POMWIG, and the solid line is the sum of the ExHuME and POMWIG predictions. 
The output of both 
Monte Carlo generators was smeared to simulate detector effects, as described in the text. The data points are the CDF Run I results \cite{cdfdijet}.}
\end{figure} 
The di-jet mass fraction $R_{JJ}$ is defined as  $R_{JJ} = M_{JJ} / M_X$, where $M_X$ is the invariant mass of the diffractive central system 
and $M_{JJ}$ is the invariant mass of the 2 highest $E_T$ jets. 
Experimentally, the total mass in the calorimeter is a difficult quantity to measure. CDF 
reconstruct $R_{JJ}$ as follows (we follow this procedure, using the smeared final state particles from the Monte Carlos)
\begin{center}
\begin{eqnarray}
M^2_{JJ} = x_1 x_2 s \\
x_1 = \frac{1}{\sqrt s} \sum_{jets 1,2} E_T e^{\eta} \\
x_2 = \frac{1}{\sqrt s} \sum_{jets 1,2} E_T e^{-\eta}
\end{eqnarray} 
\end{center}
where the $x_i$ are approximately, in the POMWIG model, the fraction of the pomeron momenta carried by the struck partons in the pomerons (often termed $\beta$).
Similarly, 
\begin{eqnarray}
\label{calo_xp}
M^2_{X} = \xi_1 \xi_2 s \\
\xi_{p} = \frac{1}{\sqrt s} \sum_{particles} E_T e^{\eta} \\
\xi_{\bar p} = \xi_{RP}
\end{eqnarray} 
where the sum is over all final state particles (excluding the outgoing protons), and $\xi_{RP}$ is the fractional longitudinal momentum loss of the 
outgoing anti-proton measured in the Roman Pot. We apply the correction factor of 1.2 to  $\xi_p$ as discussed above.

In figure 4 we show the smeared $R_{JJ}$ distributions compared to the CDF Run I results. Between $R_{JJ} = 0.2$ 
and $R_{JJ} = 0.7$, POMWIG alone gives a reasonable description of the data. This is consistent with the findings of \cite{Appleby:2001xk}. 
The POMWIG prediction is shifted to higher $R_{JJ}$ values however. There are two related reasons for this shift. Increasing the correction factor 
applied to $\xi_p$ results in a shift to lower $R_{JJ}$ (since doing so increases the measured $M_X$ value in the denominator of $R_{JJ}$). 
As we noted above, the CDF correction factor of $1.7$ had a large error due to limited statistics, 
and may therefore have been too large. Similarly, the $\xi_{\bar p}$ distribution shown in figure 2 rises more steeply with increasing $\xi_{\bar p}$ in the data 
than in POMWIG, although the statistical errors on the data are large. This again
 has the effect of increasing the $M_X$ measurement in the data, and shifting the $R_{JJ}$ distribution to lower values. 
The approach we take here is to attribute these shifts to statistical fluctuations and detector effects, which we are unable to simulate.     

For $R_{JJ} > 0.8$, POMWIG generates few events, and the distribution becomes 
dominated by exclusive events generated by ExHuME (or DPEMC). CDF had insufficient statistics to probe this region in Run I. In table 2 we show the cross sections predicted by POMWIG, ExHuME and DPEMC after detector smearing, in the range 0.01$<\xi_{p}<$0.03 using the effective gap survival of 0.27 for
 POMWIG, and the predicted gap survival factor of $\mathcal{S}^{2}$ = 0.045 for ExHuME and DPEMC.

\begin{table}[htdp]
\begin{center}
\begin{tabular}{| c | c | c | c |}
\hline
& $E_{T, min}$ (GeV)   & $\sigma_{TOT}$ (nb) & $\sigma_{R_{JJ} > 0.8}$ (nb) \\
\hline 
CDF & 7 & 43.6$\pm$4.4$\pm$21.6  & $<$ 3.7 \\
POMWIG & 7 & 42.53  & 0.01 \\ 
ExHuME & 7 &  0.59 & 0.03\\
DPEMC & 7 &  1.29 & 0.09\\
POMWIG + ExHuME & 7 & 43.12 &  0.04  \\ 
POMWIG + DPEMC & 7 & 43.8 &  0.10 \\ 

\hline
CDF & 10 & 3.4$\pm$1.0$\pm$2.0 & n/a \\
POMWIG & 10 & 6.91 &  $<$ 0.01\\ 
ExHuME & 10 & 0.28 &  0.03\\
DPEMC & 10 & 0.60 &  0.09\\
POMWIG + ExHuME & 10 & 7.19 &  0.03  \\ 
POMWIG + DPEMC & 10 & 7.52 &  0.09  \\ 
\hline
\end{tabular}
\end{center}
\caption{The cross section predictions from POMWIG (with an effective gap survival factor $S^2 = 0.27$), ExHuME and DPEMC, in the kinematic range described in the text, with detector smearing included. Also shown are the CDF Run I 
published cross sections, taken from \cite{cdfdijet}.}
\label{Run1Sigma}
\end{table}

DPEMC predicts a larger cross section than ExHuME after the CDF cuts. This is partly due to the soft $\xi$ dependence of DPEMC which leads to more events being generated in the large $\xi$ region. It should always be remembered however, that there are large uncertainties in the normalisations of the ExHuME and DPEMC predictions, and both the absolute and relative normalisations can vary within at least a factor of 2. Neither model is ruled out by Run I results. 

\subsection{CDF Run II predictions}
\label{cdf2results}
Having used the published Run I data to fix the normalisation of the POMWIG sample ($S^2 = 0.27$), we can predict $R_{JJ}$ for the increased 
statistics (and differing kinematic range) of Run II, following the approach used in the CDF preliminary central di-jet production analysis described in \cite{Gallinaro:2005qh}.
The kinematic range for Run II is as follows;
\begin{eqnarray}
E_T^{jets 1,2} > 10~{\rm GeV} \\
|\eta_{jets 1,2}| < 2.5 \\ 
0.03 < \xi_{\bar p} < 0.1 \\
\xi_p < 0.1 \\
|t_{\bar p}| < 1 {\rm GeV}^2.
\end{eqnarray}
The larger $\xi_{p}$ range is due to a new rapidity gap definition, partially imposed by the new configuration of forward detectors at CDF Run II \cite{Gallinaro:2004nd}. 
In table 3 we show the hadron level cross section predictions from POMWIG, ExHuME and DPEMC. We also show the ExHuME prediction using the CTEQ6M parton distribution functions, which lead to a larger cross section (although still significantly lower than the DPEMC prediction).

In figure 5 we show the hadron level $R_{JJ}$ distribution predicted by POMWIG, ExHuME and DPEMC. Due to the effects of parton showering and hadronisation, neither model predicts a visible excess of events at large $R_{JJ}$. In fact the majority of exclusive events lie below $R_{JJ}=0.8$.  
The DPEMC exclusive cross section is larger than that of ExHuME, as in the Run I kinematic range, although the caveats about the absolute and relative normalisations of ExHuME and DPEMC still apply.  

In figure 6 we show the $E_T$ distribution of the second highest $E_T$ jet in the `exclusive' region $R_{JJ}>0.8$, for ExHuME and  DPEMC. Here, there is a 
difference between the predictions of the models. As discussed in section 1, this is because the ExHuME (KMR) cross section falls rapidly as the di-jet $E_T$ rises, due to the increased phase space for gluon emission. In non-perturbative models, this suppression is not present, and the jet 
$E_T$ dependence of the cross section is therefore much flatter. In figure 6 we also show the jet $E_T$ dependence of events generated by ExHuME, DPEMC and POMWIG alone, normalised 
 such that all curves pass through the same point at $E_T = 10$ GeV, making the difference in the predicted slopes easier to see. In common with DPEMC, 
there is less suppression of the production cross section with increasing $E_T$ in POMWIG. Note also that the parton distribution functions used in ExHuME affect the $E_T$ distributions in a non-negligible way. CTEQ6M produces a softer $E_T$ dependence than MRST2002, although both are significantly steeper than the non-perturbative models, for the reasons discussed above.

\begin{table}[htdp]
\begin{center}
\begin{tabular}{| c | c | c | c |}
\hline
& $E_{T, min}$ (GeV)   & $\sigma_{TOT}$ (nb) & $\sigma_{R_{JJ} > 0.8}$ (nb) \\
\hline 
POMWIG & 10 & 188.16 & 0.10  \\ 
ExHuME (MRST2002) & 10 &   0.82 & 0.26 \\
ExHuME (CTEQ6M) & 10 &   1.45 & 0.43 \\
DPEMC & 10 & 2.61 & 0.80\\
POMWIG + ExHuME (MRST2002) & 10 & 188.98 & 0.36  \\ 
POMWIG + ExHuME (CTEQ6M) & 10 & 189.61 & 0.53  \\ 
POMWIG + DPEMC & 10 & 190.77 &  0.90 \\ 
\hline
POMWIG & 25 &  0.940 & 0.008 \\ 
ExHuME (MRST2002) & 25 & 0.016 & 0.012\\
ExHuME (CTEQ6M) & 25 & 0.037 & 0.027\\
DPEMC & 25 & 0.176 & 0.118\\
POMWIG + ExHuME (MRST2002) & 25 & 0.956 & 0.020\\
POMWIG + ExHuME (CTEQ6M) & 25 & 0.977 & 0.035\\
POMWIG + DPEMC & 25 & 1.116 & 0.126\\
\hline
\end{tabular}
\end{center}
\label{runIItablegen}
\caption{The cross section predictions from POMWIG (with an effective gap survival factor $S^2 = 0.27$), ExHuME and DPEMC, in the CDF Run II preliminary kinematic range as described in the text. Also shown is the ExHuME prediction using the CTEQ6M proton pdfs.} 
\end{table}%

\begin{figure}[htdb]        
\label{rjjrunIIgen}
\begin{center}
  \setlength{\unitlength}{1 mm}      
  \Large 
\begin{picture}(160,70)(0,0)
    \put(0,0){\epsfig{file=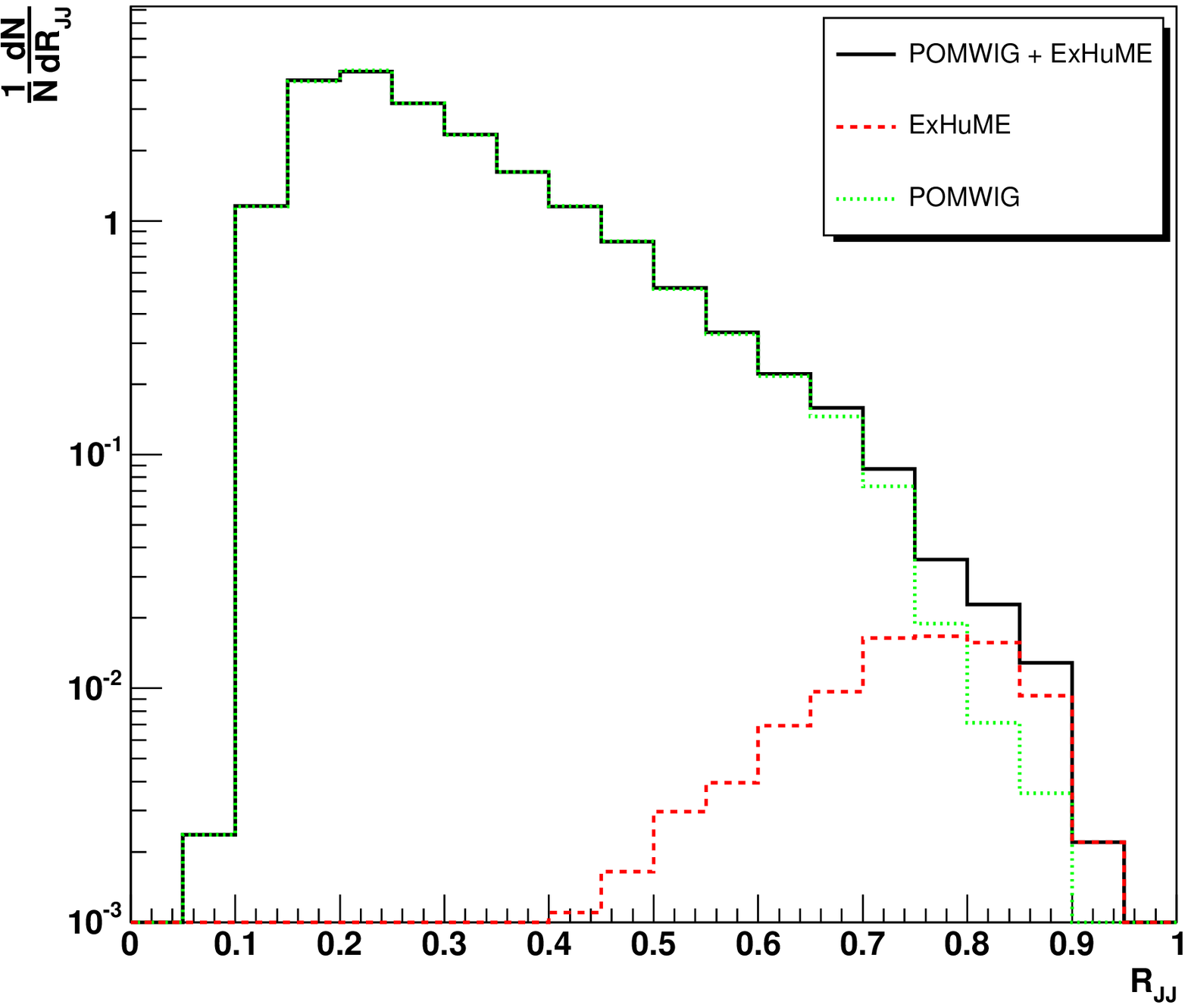,width=.5\textwidth}}
    \put(80,0){\epsfig{file=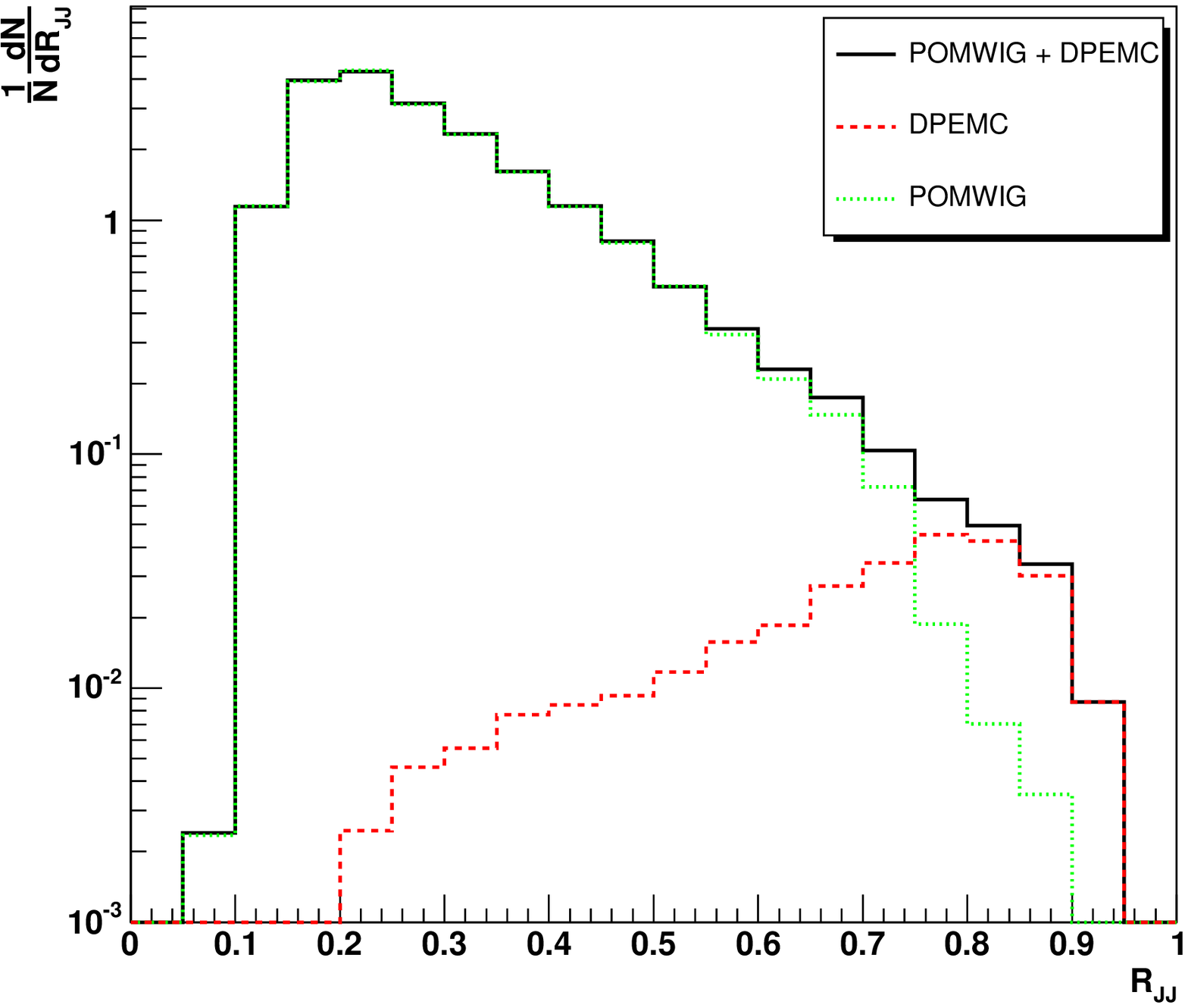,width=.5\textwidth}}
\end{picture}
\end{center}
\caption 
        {The $R_{JJ}$ distributions at the hadron level predicted by POMWIG + ExHuME (left hand plot) and POMWIG + DPEMC (right hand plot), in the CDF Run II kinematic range, as described in the text.}
\end{figure} 

\begin{figure}[htdb]        
\label{etdistrunIIgen}
\begin{center}
  \setlength{\unitlength}{1 mm}      
  \Large 
\begin{picture}(160,60)(0,0)
    \put(80,0){\epsfig{file=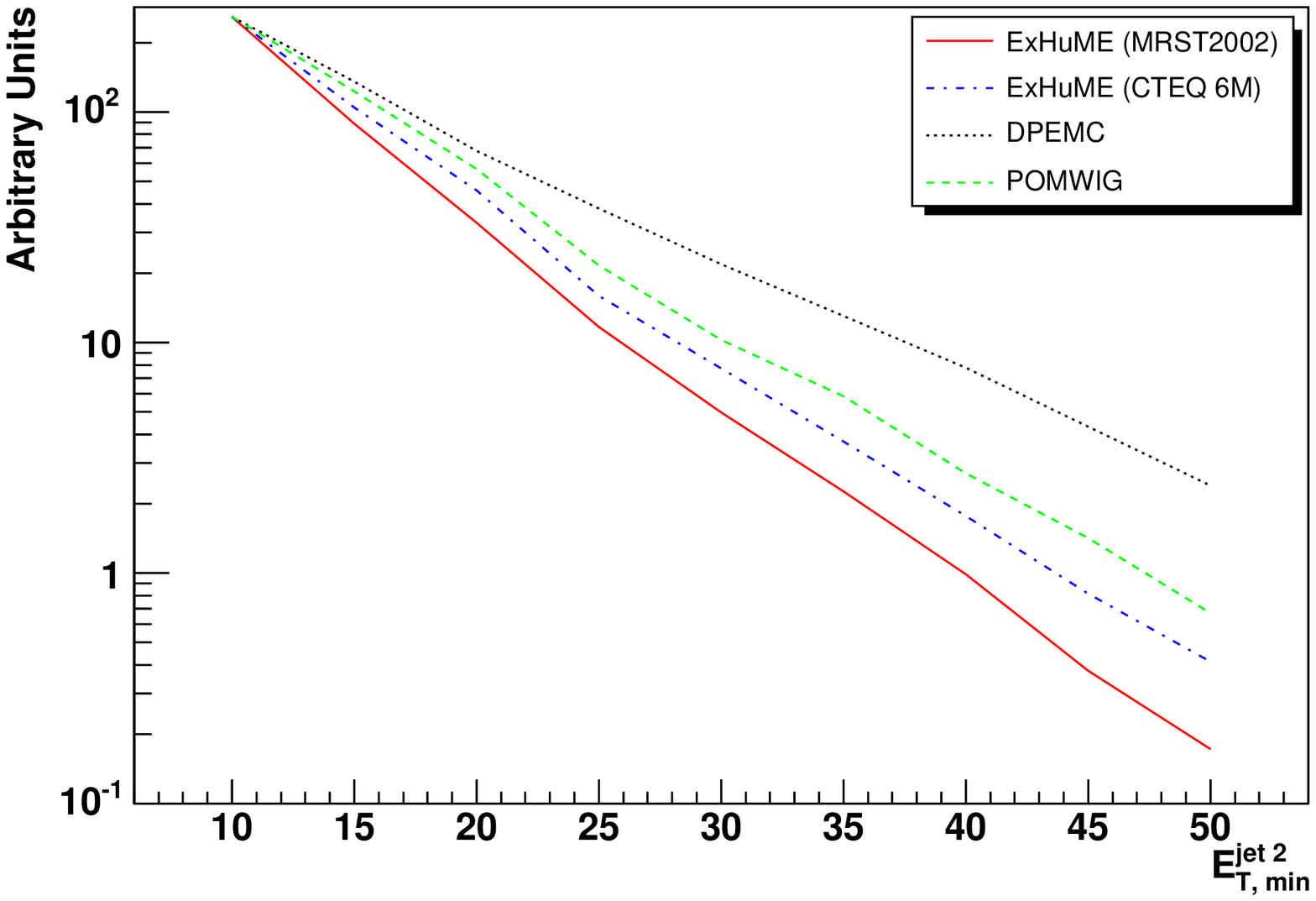,width=.5\textwidth}}
    \put(0,0){\epsfig{file=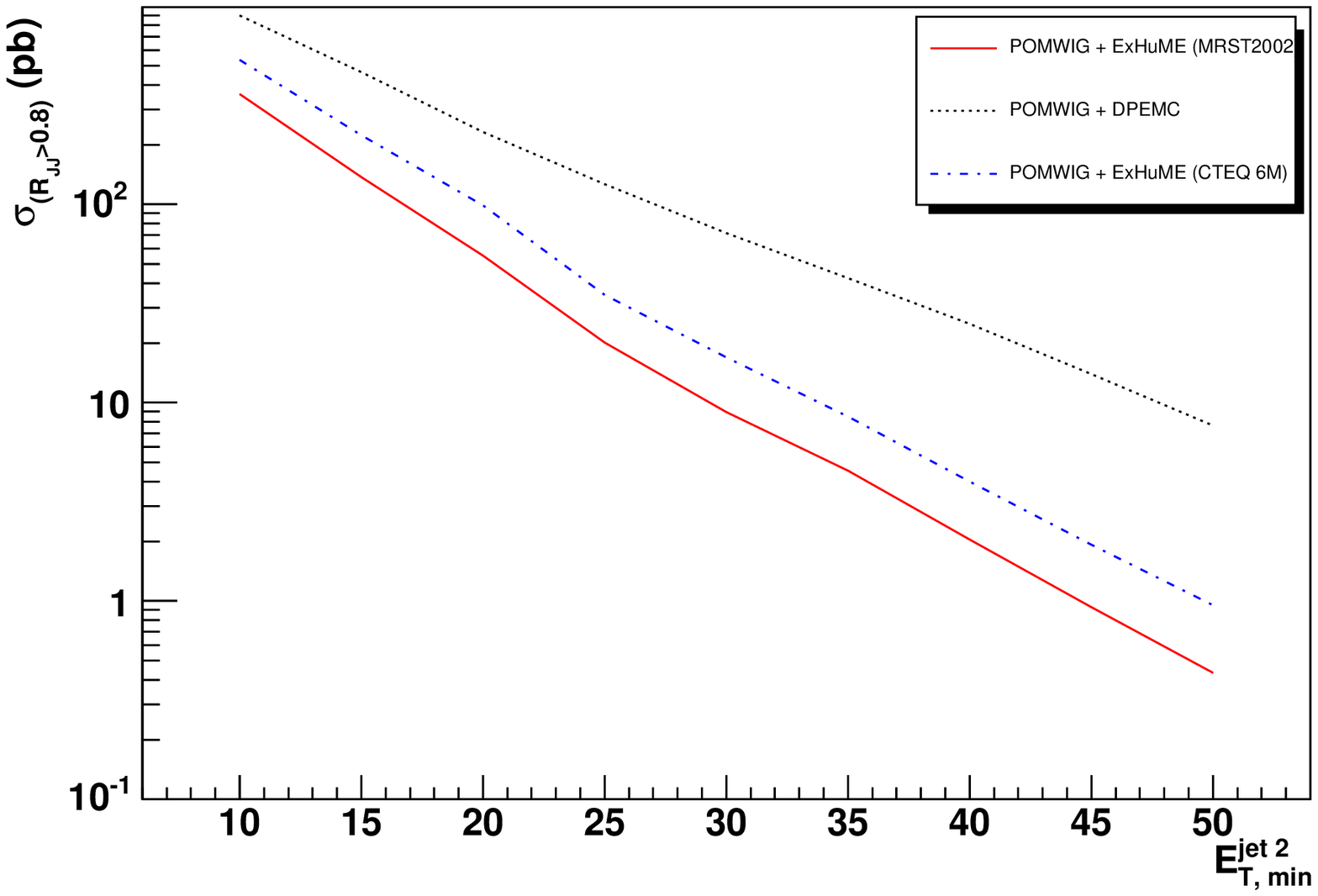,width=.5\textwidth}}
\end{picture}
\end{center}
\caption 
        {The $E_T$ distribution of the second highest $E_T$ jet in the region $R_{JJ} > 0.8$. The predictions of POMWIG + ExHuME (MRST2002), POMWIG + ExHuME (CTEQ6M) and  POMWIG + DPEMC are shown in the left hand plot. The kinematic range is as described in the text. 
In the right hand plot, the predictions of  ExHuME alone (with MRST2002 and CTEQ6M structure functions), POMWIG alone and DPEMC alone are shown with the curves 
normalised such that they all pass through the same point at $E_T = 10$ GeV.}
\end{figure} 

\begin{figure}[htdb]        
\label{etdistrunIIgen}
\begin{center}
  \setlength{\unitlength}{1 mm}      
  \Large 
\begin{picture}(160,70)(0,0)
    \put(0,0){\epsfig{file=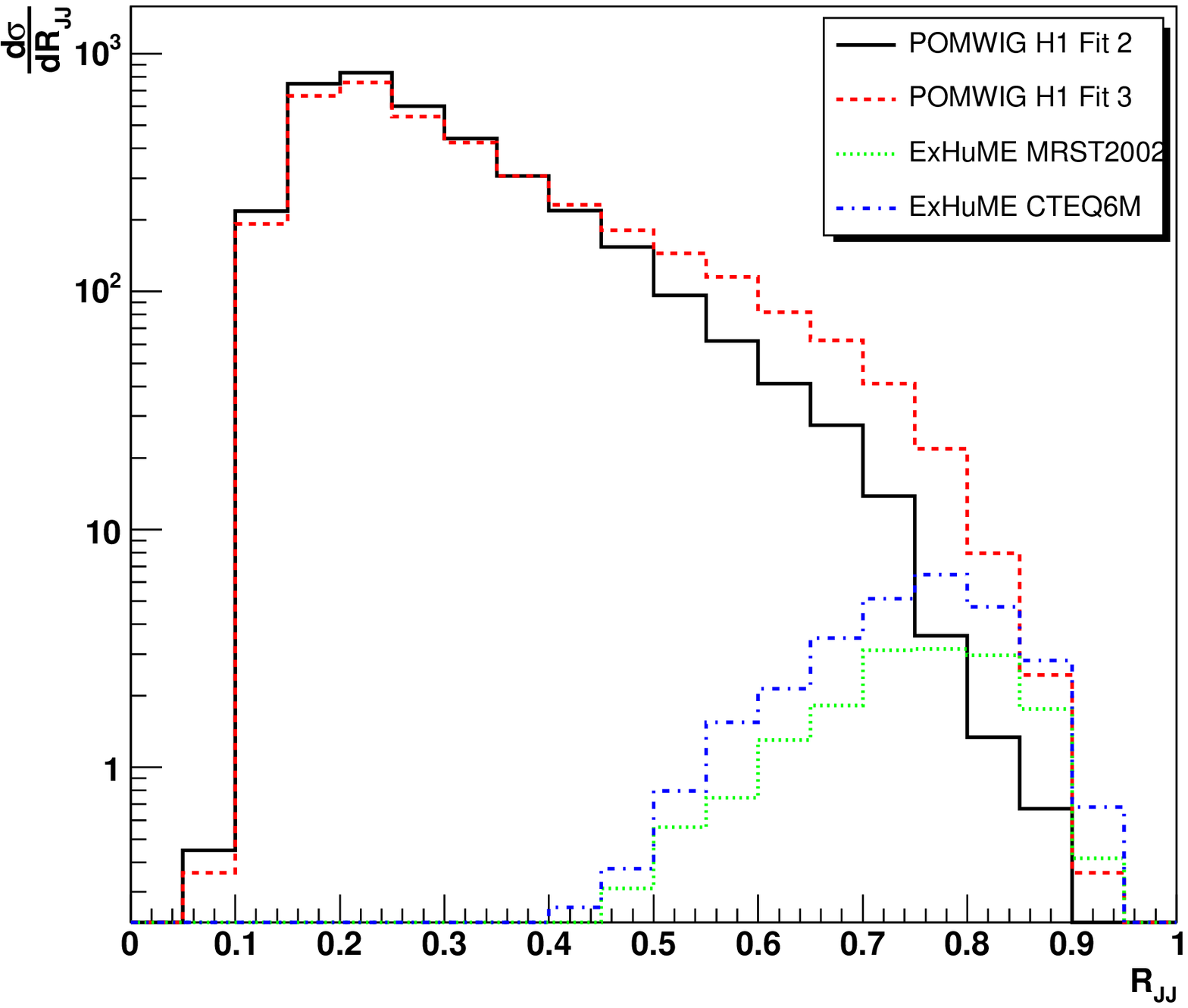,width=.5\textwidth}}
    \put(80,0){\epsfig{file=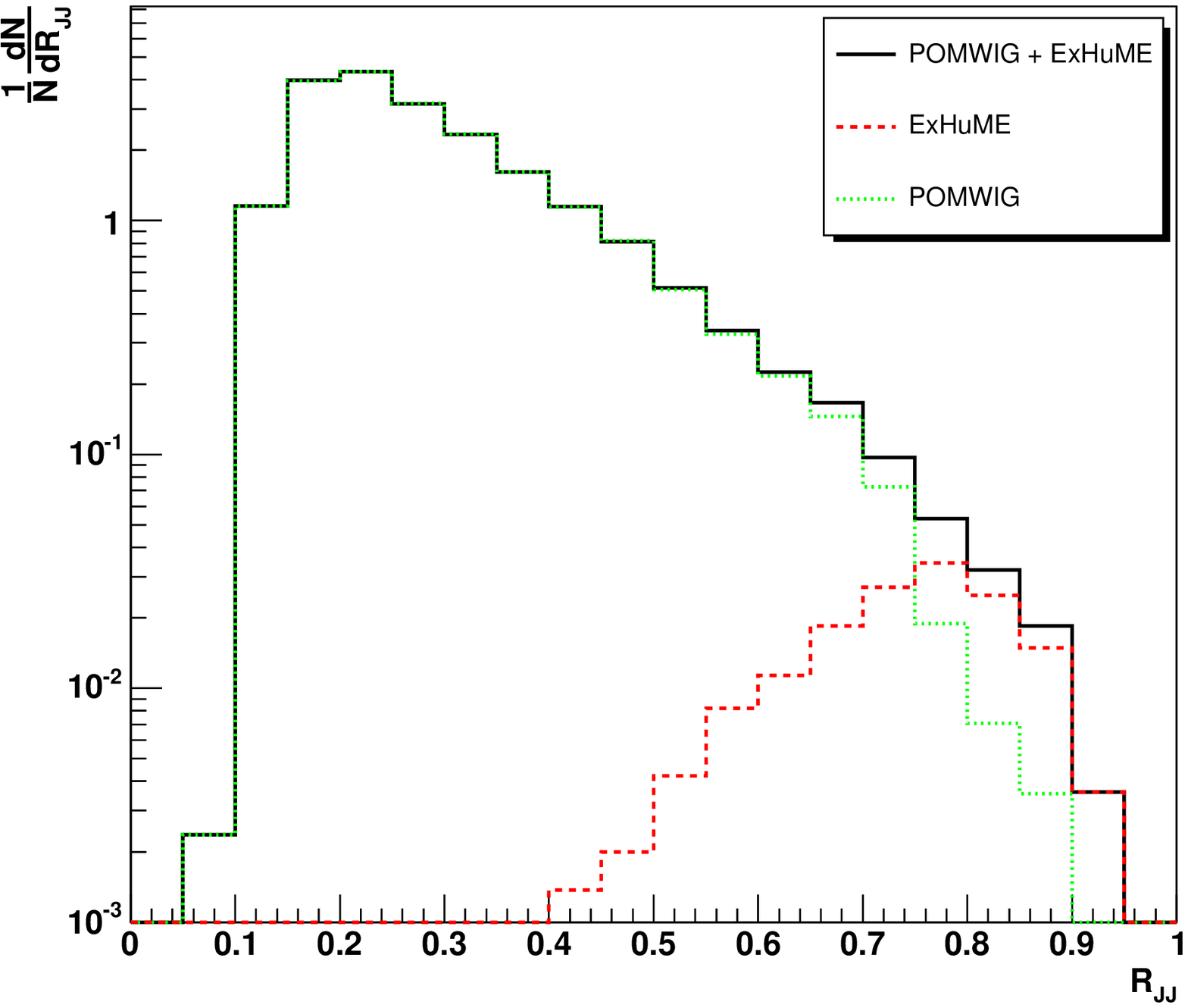,width=.5\textwidth}}
\end{picture}
\end{center}
\caption 
        {The $R_{JJ}$ distribution at the hadron level predicted by POMWIG alone, using the H1 fit 2 and H1 fit 3 'peaked gluon' diffractive structure functions, in the Run II kinematic range as described in the text. Also shown are the predictions of ExHuME alone using the MRST2002 and CTEQ6M proton structure functions (Left hand plot). The right hand plot shows the $R_{JJ}$ distribution at the hadron level predicted by POMWIG + ExHuME, where the CTEQ6M proton parton distributions are used for ExHuME.}
\end{figure} 

\begin{figure}[htdb]        
\label{etdistrunIIgen}
\begin{center}
  \setlength{\unitlength}{1 mm}      
  \Large 
\begin{picture}(160,70)(0,0)
    \put(80,0){\epsfig{file=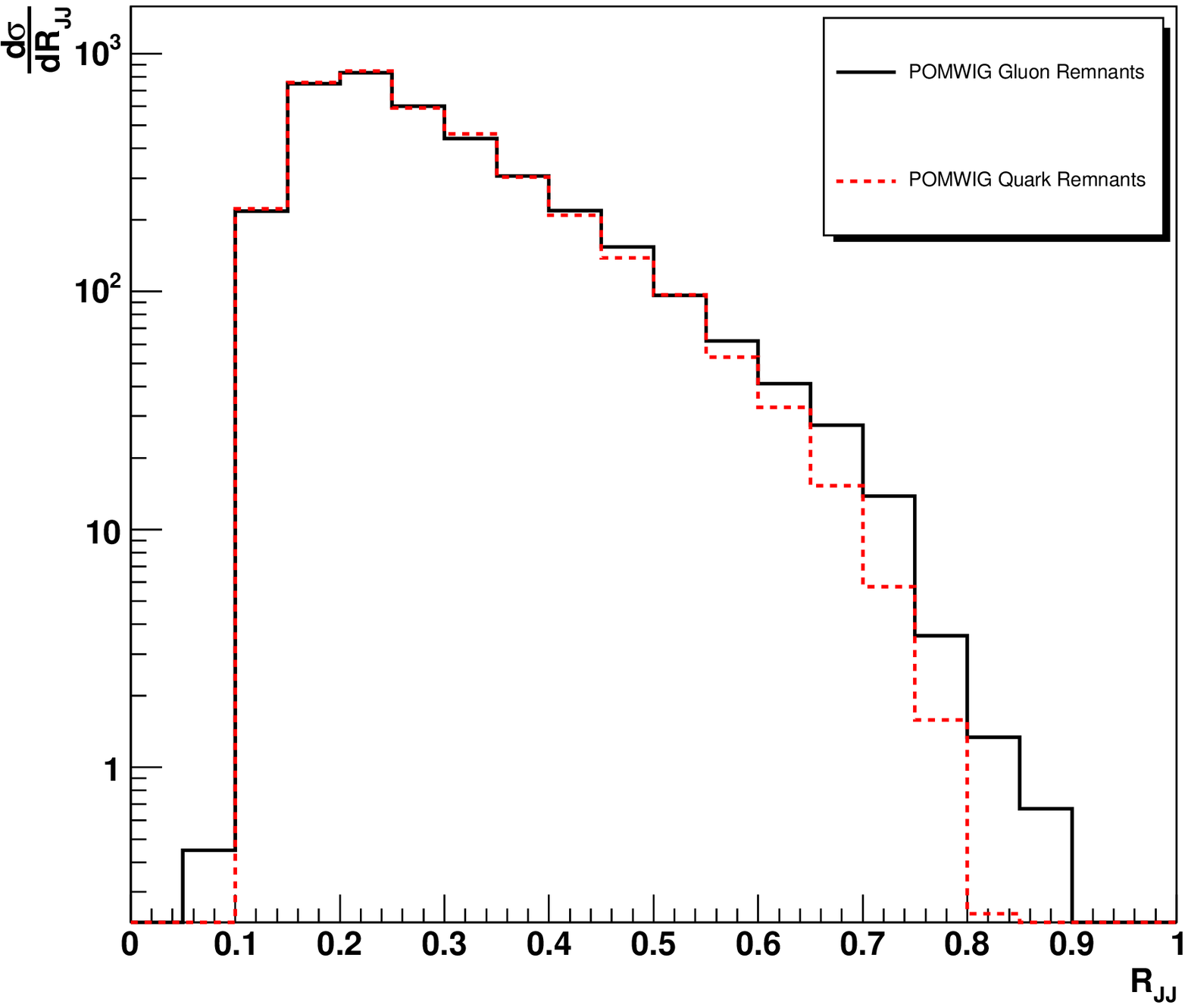,width=.5\textwidth}}
    \put(0,0){\epsfig{file=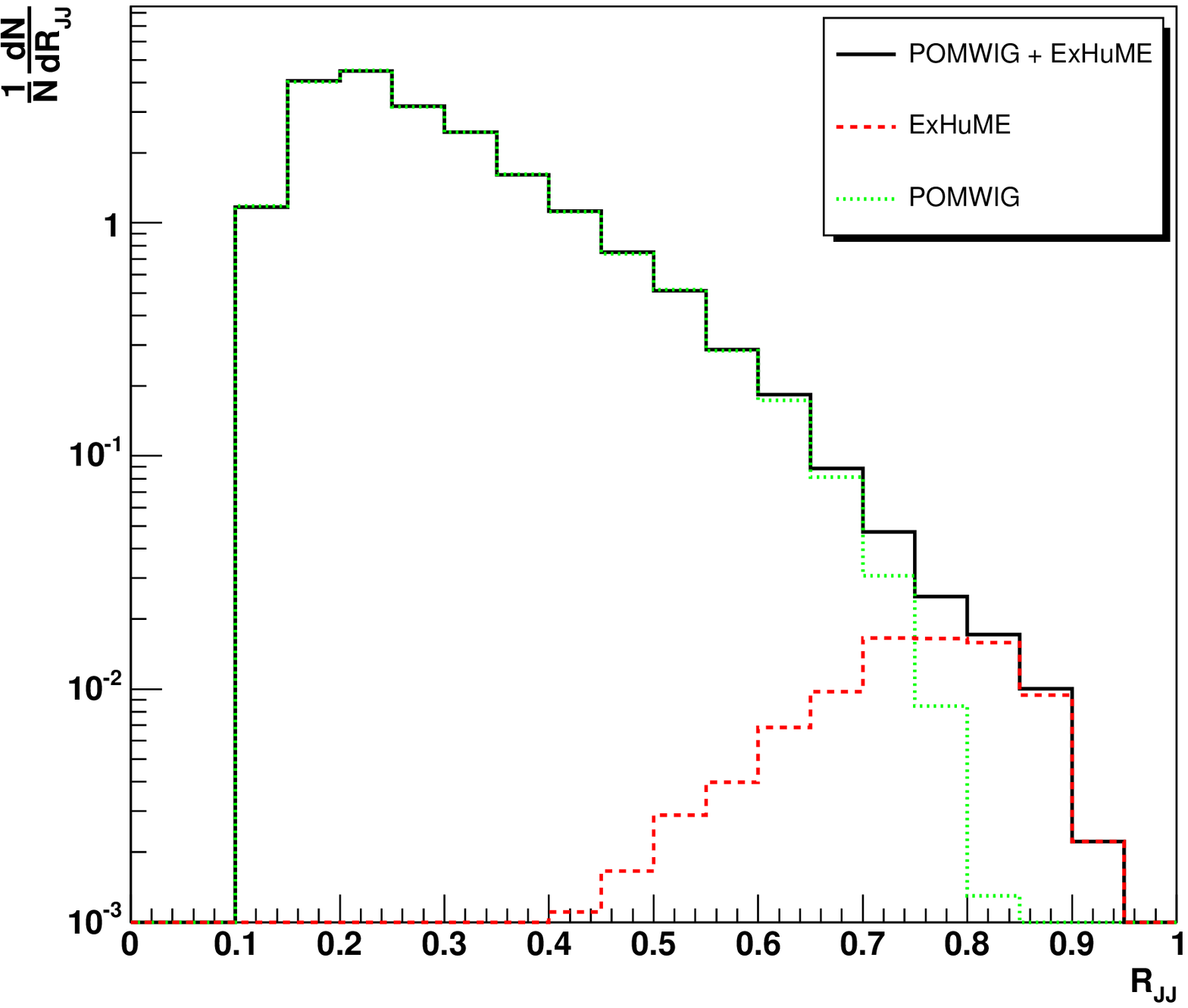,width=.5\textwidth}}
\end{picture}
\end{center}
\caption 
        {The $R_{JJ}$ distribution at the hadron level predicted by POMWIG and ExHuME, in the Run II kinematic range as described in the text, with the valence partons in the pomeron defined to be quarks (Left hand plot). The Right hand plot shows the POMWIG predictions alone with the pomeron valence partons defined to be quarks (dashed line) and gluons (solid line). In both figures, the Reggeon valence 
partons are defined to be quarks.}  
\end{figure}

\section{Summary and Discussion}
\label{summary}
We have shown that the published CDF Run 1 double pomeron exchange data do not rule out the presence of an exclusive component at the level predicted by Khoze, Martin and Ryskin (KMR). The DPEMC prediction is also not ruled out by the Run I results. 
Looking forward to the Tevatron Run II data, we have made a prediction for the di-jet mass fraction, $R_{JJ}$, based on the POMWIG Monte Carlo, which we scale to the Run I measured cross section, and the absolute prediction of 
the KMR model from ExHuME. At the time of writing, only preliminary results are available for the high - statistics Run II data. In the absence of final, published results, we present hadron 
level predictions, in the kinematic range used in the preliminary analysis \cite{Gallinaro:2005qh}. As is clear from figure 5 and table 3, POMWIG generates fewer events at 
$R_{JJ} > 0.8$ than ExHuME. With sufficient statistics, and 
if detector effects are corrected for, we might therefore expect the $R_{JJ}$ distribution in the data to be higher than 
the POMWIG prediction at $R_{JJ} > 0.8$. This could be interpreted as a sign of an exclusive component in the data, given 
that POMWIG is able to adequately model the high $R_{JJ}$ region. This is an assumption, but we believe it to be a reasonable one, for the following reason. In order to produce more events 
at large $R_{JJ}$, the pomeron structure function would have to be larger at high-$\beta$ than it is in POMWIG. In figure 7, we compare the effect of varying the structure function by comparing 
the POMWIG default (H1 fit 2 \cite{Adloff:1997sc}) to the H1 fit 3 peaked gluon. Even with the relatively extreme high-$\beta$ behaviour of the H1 fit 3 gluon distribution (which is certainly too large), 
an excess at high $R_{JJ}$ would still be present in the data above the POMWIG prediction if an exclusive component at the level predicted by KMR is present, since the cross section from ExHuME alone is as large or larger than the POMWIG H1 fit 3 prediction in the highest $R_{JJ}$ bins. As discussed above, there is an uncertainty in the absolute magnitude of the KMR prediction, which is largely due to the uncertainty in the knowledge of the gluon distributions of the proton.
In figure 7 we also show the ExHuME prediction using the CTEQ6M parton distributions (see also table 3). In this case, the excess over the POMWIG prediction is larger.    

The treatment of the pomeron remnant can also have an effect on the POMWIG predictions at high $R_{JJ}$. As discussed in section 2, POMWIG v1.3 introduces the option of defining the pomeron valence partons to be either quarks or gluons. Since in the H1 picture, the pomeron is predominantly a gluonic object, the more natural choice is to define the valence partons to be gluons. In figure 8 we compare the $R_{JJ}$ distributions when the pomeron valence partons are defined to be quarks (the POMWIG default behaviour prior to v1.3) or gluons. The POMWIG 
prediction is significantly lower at high $R_{JJ}$ for the quark case. 
This is because in di-jet production, the parton entering the hard scatter from the pomeron is more likely to be a gluon. If the valence partons in the Pomeron are defined to be quarks, HERWIG will be forced to continue parton showering until a quark-antiquark pair is produced. This results in increased activity in the remnant region, and hence high $R_{JJ}$ values are less likely.

Both the DPEMC and POMWIG models have significantly different jet $E_T$ dependences to ExHuME (with both the CTEQ6M and MRST2002 partons distributions), as shown in figure 6. The steeper jet $E_T$ dependence seen in ExHuME is a direct result of the fact 
that the KMR model takes account of the increasing phase space for gluon emission as $E_T$ increases. If, on the other hand, the high $R_{JJ}$ events are due to the tail of a POMWIG (or a DPEMC) - like model, 
then a flatter $E_T$ dependence should be observed. The jet $E_T$ distribution in the high $R_{JJ}$ region is therefore a possible test for the presence of a perturbative exclusive component (i.e. KMR) in the 
double pomeron di-jet data.   

\section*{Acknowledgments}
We would like to thank Maarten Boonekamp, Jeff Forshaw, Michele Gallinaro, Dino Goulianos, Valery Khoze and Koji Terashi for useful discussions, suggestions and encouragement throughout the project.
This work was funded in the UK by PPARC.

\end{document}